\documentclass[5p,twocolumn]{elsarticle}
\usepackage[T1]{fontenc}
\usepackage[utf8]{inputenc}
\usepackage{lmodern}
\usepackage{epsfig} 
\usepackage{graphicx}
\usepackage{amsmath}
\usepackage{color}
\usepackage{slashed}
\usepackage{subfig}
\usepackage{mathtools}
\usepackage{natbib}

\biboptions{sort&compress}

\begin{document}

\begin{frontmatter}

\title{Vacuum energy with mass generation and Higgs bosons}
\author[1,2]{Steven D. Bass}
\ead{Steven.Bass@cern.ch}
\address[1]{Kitzb\"uhel Centre for Physics,
Kitzb\"uhel, Austria}
\address[2]{
Marian Smoluchowski Institute of Physics, Jagiellonian University, 
PL 30-348 Krakow, Poland}
\author[3]{Janina Krzysiak}
\ead{Janina.Krzysiak@ifj.edu.pl}
\address[3]{Institute of Nuclear Physics,
Polish Academy of Sciences,
ul. Radzikowskiego 152,
PL 31-342 Krakow, Poland}

\begin{abstract}
We discuss the Higgs mass and cosmological constant hierarchy
puzzles with emphasis on the interplay of Poincare invariance, mass generation and renormalization group invariance.  
A plausible explanation involves an emergent Standard Model with the cosmological constant scale 
suppressed by power of the large scale of emergence. 
In this scenario the cosmological constant scale and neutrino masses should be of similar size.
\end{abstract}


 \end{frontmatter}

\section{Introduction}

Particle Physics comes with two mysterious hierarchies of scales.
The mass of the Higgs boson discovered at CERN is about 125 GeV~\cite{Aad:2012tfa,Chatrchyan:2012xdj}.
Theoretically the Higgs mass squared comes with a quadratically divergent counterterm which naively would push its value towards the Planck scale.
The cosmological constant or vacuum energy density which drives the accelerating expansion of the Universe is characterized by a scale 0.002 eV~\cite{Aghanim:2018eyx}, 
very much less than the Higgs mass, QCD and Planck scales.
It is an open question whether these two puzzles might be connected.
Here we discuss 
these scale hierarchies
with focus on the interplay of Poincare invariance, mass generation and renormalization group invariance. 
We argue that a plausible explanation involves an emergent Standard Model 
\cite{Jegerlehner:2013cta,Bass:2020gpp}
and the measured cosmological constant
scale
associated with higher dimensional 
terms in the action,
suppressed by power of the large emergence scale. 
In this scenario it is natural that the cosmological constant scale and neutrino masses should be of similar size.

In Section 2 we next discuss 
the scale hierarchies associated with ultraviolet divergences 
and renormalization, 
e.g. the Higgs mass counterterm and zero-point energies induced by quantization.
Both the Higgs mass counterterm and 
net zero-point energy contribution to the cosmological constant
are related to the relative contributions of bosons and fermions, including possible extra Higgs bosons. 
Consequences for new particles at LHC energies are discussed.
Section 3 discusses the effect of running masses and couplings,
how the Standard Model parameters are linked to physics in the
ultraviolet.
Finally, in Section 4 we consider the cosmological constant and
explain why a tiny non-zero value fits with possible emergent gauge symmetry.
Large zero-point energy contributions to the cosmological constant are renormalization scale dependent and decouple.
With emergence,
contributions from potentials linked to the Higgs and
QCD condensates are cancelled 
at mass dimension four in the action.

\section{Scale hierarchies and renormalization}

In the Standard Model the masses of the W and Z gauge bosons and charged fermions come from coupling to the Higgs boson with a finite Higgs vacuum expectation value, vev. 
The renormalized Higgs mass squared comes with the divergent counterterm
\begin{equation}
m_{h \ {\rm bare}}^2 
= m_{h \ {\rm ren}}^2 + \delta m_h^2
\end{equation}
where
\begin{equation}
\delta m_h^2 
=
\frac{K^2}{16 \pi^2}
\frac{6}{v^2} 
\biggl(
m_h^2 + m_Z^2 + 2 m_W^2 - 4 m_t^2
\biggr)
\end{equation}
relates the renormalized and bare Higgs mass.
Here $K$ is an ultraviolet 
scale characterizing the limit to where the Standard Model should work, $v$ is the Higgs vev and the $m_i$ are the Higgs, Z, W and top quark masses. We neglect contributions from lighter mass quarks.

Boson and fermion contributions enter with different signs and come with different renormalization scale dependence. The renormalized and bare masses would coincide with no hierarchy puzzle if 
\begin{equation}
2 m_W^2 + m_Z^2 + m_h^2 = 4 m_t^2 .
\end{equation}
This equation is the Veltman condition \cite{Veltman:1980mj}.
It implies a collective cancellation between bosons and fermions. Taking the pole masses for the W, Z and top quark (80, 91 and 173 GeV) would require a Higgs mass of 314 GeV, much above the measured value. Running masses are discussed below.

A similar discussion follows for the vacuum
zero-point energies,
ZPEs, of quantum field theory which are induced 
by quantization \cite{Bjorken:1965zz}
and important in the 
cosmological constant~\cite{Weinberg:1988cp}
together with potentials associated with vacuum condensates.
We work in flat space-time. 
Zero-point energies come with ultraviolet divergence requiring regularization and renormalization,
\begin{equation}
\rho_{\rm zpe} =
\frac{1}{2} \sum \{\hbar\omega \}
=
\frac{1}{2} \hbar
\sum_{\rm particles} g_i \int_0^{k_{\rm max}}
\frac{d^3 k}{(2 \pi)^3} \sqrt{k^2 + m^2}
.
\end{equation}
Here $\frac{1}{2} \{ \hbar \omega \}$ 
denotes the eigenvalues of the free Hamiltonian and
$\omega = \sqrt{k^2 + m^2}$
where $k$ is the wavenumber and $m$ is the particle mass;
$g_i = (-1)^{2j} (2j+1) f$
is the degeneracy factor for a particle $i$ of spin $j$, 
with $g_i  >0$ for bosons and $g_i < 0$ for fermions.
The minus sign follows from the Pauli exclusion principle and 
the anti-commutator relations for fermions.
The factor $f$ is 1 for bosons, 
2 for each charged lepton
and 6 for each flavour of quark
(2 charge factors for the quark and antiquark, 
 each with 3 colours).

It is important to choose a Lorentz covariant 
regularization 
to ensure that the renormalized zero-point energy satisfies 
the correct vacuum equation of state. 
Dimensional regularization with minimal subtraction, 
$\overline{\rm MS}$, is a good regularization.
One finds
\begin{equation}
\rho_{\rm zpe} = - p_{\rm zpe}
=
- 
\hbar \ g_i \
\frac{m^4}{64 \pi^2} 
\biggl[ \frac{2}{\epsilon} + \frac{3}{2} - \gamma
- \ln \biggl( \frac{m^2}{4 \pi \mu^2} \biggr) \biggr]
+ ...
\end{equation}
from particles with mass $m$~\cite{Martin:2012bt}.
Here 
$p_{\rm zpe}$ is the pressure,
$D=4-\epsilon$ the number of dimensions, 
$\mu$ the renormalization scale and $\gamma$ is Euler's constant.
If one instead uses a brute force cut-off on the 
divergent integral, 
the leading term in the ZPE proportional 
to $k_{\rm max}^4$ 
obeys the radiation equation of state $\rho = p/3$.

Equation (5) means that the ZPE vanishes for massless particles,
e.g. photons. 
For Standard Model particles it
is induced by the Higgs mechanism.
Boson and fermion contributions to the net zero-point energy 
come with different signs. 
This led Pauli to suggest a collective cancellation of the ZPE \cite{Pauli:1971wp}, much like the Veltman condition for the 
Higgs mass squared. 
If we take the ZPEs as physical, then this Pauli constraint 
for cancelling the ZPE gives new constraints on possible extra particles
\cite{Pauli:1971wp,Kamenshchik:2018ttr}
\begin{eqnarray}
\sum_i g_i m_i^4 &=& 0
\nonumber \\
\sum_i g_i m_i^4 \ln m_i^2 &=& 0 .
\end{eqnarray}
For the Standard Model
with the physical W, Z and 
top-quark masses, these two equations would need a 
Higgs mass of about 319 GeV and 311 GeV respectively, 
close to the Veltman value of 314 GeV.
With the Standard Model particle masses, 
the ZPE is negative and fermions dominated.
We need some extra bosons if we want this to work.
(Contributions to the total vacuum energy 
from the Higgs potential and QCD condensates
are discussed in Section 4 below.)

Obvious first candidates are 
2 Higgs Doublet Models~\cite{Ivanov:2017dad}.
These are a simple extension of the Standard Model.
One introduces a second Higgs doublet.
There are 5 Higgs bosons, two neutral scalars $h$ and $H$, 
one pseudoscalar $A$
and two charged Higgs states $H^{\pm}$.
Since the 125 GeV Higgs-like scalar discovered at CERN 
in 2012~\cite{Aad:2012tfa,Chatrchyan:2012xdj}
has so far showed no departure from Standard Model predictions, 
it must be assumed 
in any model with extra Higgs states
that 
one of the neutral scalars $h$ 
is a lot like the Standard Model Higgs.

The possible extra Higgs states are looked 
for in direct searches \cite{Atlas:note,Flechl:2019dtr}.
The parameter space is constrained 
with lower bounds on the masses
from
global electroweak fits \cite{Haller:2018nnx} 
and rare B-decay processes 
\cite{
Misiak:2017bgg,Arbey:2017gmh}.
Different model scenarios depend on the fermion to Higgs 
couplings.
The most constrained are Type II models with 
600 GeV $< m_{H^{\pm}}$,
530 GeV $< m_A$
and 400 GeV $<m_H$.
Here one doublet couples to up type quarks and one to down type quarks and leptons. 
Others are the type I fermiophobic model where all fermions couple to just one doublet, lepton specific (one doublet to quarks and one to leptons) and flipped 
(same as type II except leptons couple to the doublet 
 with up type quarks).
There are also inert models where only one doublet acquires a 
vev and couples to fermions.
These models are less well constrained.
For the Veltman condition extended to 2HDMs, 
a favoured benchmark point
is quoted in the Type II model 
with $m_H \sim 830$ GeV and $m_A, m_{H^{\pm}} \sim 650$ GeV 
\cite{Darvishi:2017bhf}.
Within the mass constraints quoted for the Type II models, 
bosons win! 
We would need also extra fermions in the energy range of 
the LHC to cancel the Pauli condition with the allowed masses.

\section{Running masses}

We next turn to running masses.

Both the 
Veltman and Pauli constraints are evaluated from 
loop diagrams so the masses which appear there 
are really
renormalization group, RG, scale dependent.
Boson and fermion contributions enter with different signs and evolve differently under RG evolution
which means they have a chance 
to cross zero deep in the ultraviolet.
With the particle masses and couplings measured at the LHC, 
the Standard Model works as a consistent theory up to the 
Planck scale.
One finds that the electroweak vacuum sits very close to the border of stable and metastable suggesting possible new critical phenomena in the ultraviolet, within 1.3 standard deviations of being stable on relating the top quark 
Monte-Carlo and pole masses if we take just 
the Standard Model with no coupling 
to undiscovered new particles \cite{Bednyakov:2015sca}. 
The question of vacuum stability depends on whether 
the Higgs self-coupling crosses zero or not deep in 
the ultraviolet
and involves a delicate balance of Standard Model parameters.
The Higgs and other particle masses 
might be determined by physics close to the Planck scale.

The scale of Veltman crossing is calculation dependent.
Values reported are
$10^{16}$ GeV with a stable vacuum
\cite{Jegerlehner:2013cta}, 
about $10^{20}$ GeV
\cite{Masina:2013wja} 
and much above 
the Planck scale of $1.2 \times 10^{19}$ GeV
\cite{Degrassi:2012ry,Hamada:2012bp}
with a metastable vacuum.
With the Standard Model evolution code~\cite{Kniehl:2016enc}
crossing is found at the Planck scale with a Higgs mass about
150 GeV, and not below with the measured mass of 125 GeV.
(The 125 GeV mass is close to the minimum needed 
 for vacuum stability.)

\section{The cosmological constant}

Experimentally, vacuum energy becomes 
important through the cosmological constant $\Lambda$. 
This measures the vacuum energy density
$\rho_{\rm vac} = \Lambda / (8 \pi G)$ where $G$ is Newton's
constant.
The cosmological constant appears 
on the right-hand side of 
Einstein's equations of General Relativity
\begin{equation}
R_{\mu \nu} - \frac{1}{2} g_{\mu \nu} \ R = 
- \frac{8 \pi G}{c^2} T_{\mu \nu} + \Lambda g_{\mu \nu} .
\end{equation}
Here $R_{\mu \nu}$ is the Ricci tensor, 
$R$ is the Ricci scalar
and 
$T_{\mu \nu}$ is the energy-momentum tensor 
for excitations above the vacuum.
The cosmological constant
receives contributions from 
the ZPEs,
any (dynamically generated) potential in the vacuum,
e.g. associated with the Higgs and QCD condensates, 
and a renormalized version of 
the bare gravitational term $\rho_{\Lambda}$
~\cite{Weinberg:1988cp,Sola:2013gha}.

The net vacuum energy density after turning on particle interactions,
\begin{equation}
\rho_{\rm vac} = \rho_{\rm zpe} + \rho_{\rm potential} 
+ \rho_{\rm \Lambda},
\end{equation}
is renormalization scale invariant.
It drives the accelerating expansion of the Universe and is
independent of how a theoretician might choose to calculate it,
\begin{equation}
\frac{d}{d \mu^2} \rho_{\rm vac} =0 . 
\end{equation}

Before we couple to gravity only energy differences have 
physical meaning,
which then allows us to 
cancel the zero-point energy contribution
by normal ordering
before consideration of vacuum potentials 
induced by spontaneous symmetry breaking.
The Casimir force which is sometimes claimed as experimental evidence for ZPEs
can also be calculated without reference to ZPEs
\cite{Jaffe:2005vp}.
Calculation of the Casimir effect involves 
Feynman graphs with external lines whereas the ZPE does not, 
just closed loops.

The ZPE contributions in Eq.(5) are scale dependent
both through explicit $\mu^2$ dependence and through
the running masses.
For QCD, the degrees of freedom depend on the resolution.
Deep in the ultraviolet 
one has asymptotic freedom.
For massless quarks, the ZPE vanishes.
Quark-gluon interactions are chiral symmetric at these
scales. 
In the infrared 
confinement and dynamical chiral symmetry breaking take over:
the degrees of freedom are protons, neutrons,
pions, nucleon resonances...
If energy conservation held for the ZPE (plus QCD potential
terms) one would find a constraint condition on the hadron
spectrum from summing over hadronic ZPE contributions.
The Higgs potential is RG scale dependent through 
the scale dependence of the Higgs mass and Higgs self-coupling,
which determines the stability of the electroweak vacuum.
QCD quark and gluon condensates also enter.
Renormalization scale dependence cancels to give the scale
invariant $\rho_{\rm vac}$.
The important question is whether,
after we sum over the ZPEs,
potentials in the vacuum
and the $\rho_{\Lambda}$ contribution in Eq.~(8),
there is anything left over.
How big is the remaining $\rho_{\rm vac}$?

A clue may be the curious 
result that with a finite cosmological constant
Einstein's equations 
have no solution 
where $g_{\mu \nu}$ is the constant Minkowski metric \cite{Weinberg:1988cp}
(A non-vanishing
$\rho_{\rm vac}$ acts as a gravitational source 
 which generates a dynamical space-time, with accelerating
 expansion for positive $\rho_{\rm vac}$).
For the vacuum with net constant field,
$\rho_{\rm vac} \neq 0$,
space-time
translational invariance is broken without extra fine tuning. 
This, in turn, challenges the flat space-time with covariance
assumed in Eq.(5).

So far we have treated the Standard Model as fundamental.
Interactions are determined by gauge symmetries.
The theory is covariant and renormalizable 
and
described by an action with terms of dimension four or less.

The Standard Model and its gauge symmetries 
might be emergent
\cite{Jegerlehner:2013cta,Jegerlehner:1998kt,Bjorken:2001pe,Forster:1980dg,Witten:2017hdv,Bass:2020gpp}.
For example, 
Standard Model particles 
could be the long-range, collective excitations of a statistical system near to its critical point
that resides close to 
the Planck scale~\cite{Jegerlehner:2013cta}.
With the Standard Model as an effective theory emerging 
in the infrared,
low-energy global symmetries can 
be broken through additional higher dimensional terms,
suppressed by powers of a 
large ultraviolet mass scale~\cite{Witten:2017hdv,Bass:2020gpp}. 
Gauge symmetries would be exact, 
modulo the Higgs coupling, within the effective theory.
Suppose the vacuum including condensates with finite vevs
is translational invariant and 
flat space-time is consistent at dimension four, 
just as suggested by the success of the Standard Model.
Then 
the RG invariant scales
$\Lambda_{\rm qcd}$ and 
electroweak $\Lambda_{\rm ew}$
might enter the cosmological constant
with the scale of the leading term
suppressed by 
$\Lambda_{\rm ew}/M$, where $M$ is the scale of emergence
(that is, $\rho_{\rm vac} \sim (\Lambda_{\rm ew}^2/M)^4$
 with one factor of 
 $\Lambda_{\rm ew}^2/M$ for each dimension of space-time). 
This scenario, if manifest in nature, 
would explain why the cosmological constant scale 0.002 eV 
is similar 
to what we expect for the neutrino masses 
\cite{Altarelli:2004cp},
themselves linked to a dimension five operator with
$m_{\nu} \sim \Lambda_{\rm ew}^2/M$ 
and Majorana neutrinos 
\cite{Weinberg:1979sa}.
The cosmological constant would vanish at dimension four.
That is,
$\rho_{\rm vac}=0$ 
follows as a renormalization condition at dimension four 
set by space-time translational invariance,
even in the presence of Higgs and QCD vacuum condensates.
The precision of global symmetries in our experiments,
e.g. 
lepton and baryon number conservation, 
tells us that the scale of emergence should 
be deep in the ultraviolet,
much above the Higgs and other Standard Model particle masses.

\bibliographystyle{unsrt}

\end{document}